%% file: main.tex
\documentclass[runningheads]{llncs}
\usepackage[T1]{fontenc}
\usepackage{graphicx}
\usepackage{breakurl}
\usepackage[hyphens]{url}

\usepackage{amsmath,amsfonts,amssymb}
\usepackage{multirow}

\usepackage{calc}

\usepackage{color}

\usepackage{esvect}
\usepackage{enumitem}
\usepackage{macros}

\newcommand{\rotR}{\mathord{\rotatebox[origin=c]{180}{$\mathsf{R}$}}}

\begin{document}

\title{An ASP-based approach to Solving General Stochastic Two-Player Games}
%
%

%
\author{Yifan He \and Michael Thielscher}
\authorrunning{Y. He \and M. Thielscher}
%
\institute{The University of New South Wales, Sydney, Australia\\
\email{\{yifan.he1,mit\}@unsw.edu.au}}
\maketitle              
\begin{abstract}
The Game Description Language (GDL) is a widely used formalism for specifying general games. Due to their similar syntax and semantics, Answer Set Programming (ASP) and its extensions have been applied to single- and two-player deterministic turn-taking GDL games. This paper presents the first ASP-based approach for solving two-player turn-taking GDL games with uncertainty. We introduce Stochastic Answer Set Programming (SQASP) to encode the maximally achievable winning probability for a given player in stochastic GDL games, and develop a translation-based solver that evaluates SQASP programs by converting them to Extended Stochastic Satisfiability. Our empirical results show that the proposed approach is competitive with forward search on small stochastic games and can potentially support general game players in endgame evaluation.
\keywords{General Game Playing  \and Answer Set Programming \and Extended Stochastic Satisfiability}
\end{abstract}
\input{intro}

\input{background}
\input{encoding}

\input{experiment}
\input{conclusion}
\bibliographystyle{splncs04}
\bibliography{main}
\end{document}

%% file: intro.tex
\section{Introduction}
General game players are systems that play strategic games based on descriptions provided at runtime, without prior knowledge of the game rules. The Game Description Language (GDL) is a widely used formalism for specifying such games, and its extension GDL-II supports randomness and partial observability~\cite{genesereth2014general}. In GDL-II, randomness is modeled by a special player $\gdl{random}$ that selects legal actions uniformly. A variety of approaches to \emph{play\/} general games have been proposed, most notably Minimax with automatically generated heuristics~\cite{schiffel2007fluxplayer}, Monte Carlo tree search~\cite{bjornsson2009cadiaplayer}, constraint satisfaction~\cite{koriche2016general}, and deep reinforcement learning~\cite{goldwaser2020deep}, whereas relatively few attempts were made in game \emph{solving\/}. While solving all GDL-II games is challenging, it is nevertheless interesting to identify classes of games that admit perfect play.

Due to their similar syntax and semantics, Answer Set Programming (ASP) and its extensions~\cite{fandinno2021planning,gebser2012answer} have been applied to solving single- and two-player deterministic games with perfect information~\cite{he2024solving,thielscher2009answer}. Although in both cases, the efficiency of the ASP-based game solver is comparable to forward search, neither solver supports GDL games with randomness or partial observability.

In this paper, we address the first limitation by presenting an ASP-based approach for solving two-player, stochastic, turn-taking GDL games. Such games contain a player, an opponent, and a stochastic environment. Since existing ASP frameworks cannot express both adversarial and stochastic behavior, we introduce a new language, \emph{Stochastic Answer Set Programming (SQASP)}, and show how it can be used to determine the maximally achievable winning probability for a given player in stochastic GDL games. We further develop a translation-based SQASP solver that evaluates SQASP programs by reducing them to \emph{Extended Stochastic Satisfiability (XSSAT)}~\cite{teige2012stochastic}, an extension of Quantified Boolean Formulas (QBF) with chance quantifiers, and invoking an XSSAT solver~\cite{fan2023sharpssat}. We also study the impact of \emph{quantifier shifting\/}, which is an effective encoding technique for deterministic games~\cite{he2024solving}, in the stochastic setting. Our empirical evaluation with stochastic variations of the standard GDL games Tic-Tac-Toe, \mbox{Connect-4}, and Nim demonstrates that the proposed approach is comparable to forward search and can potentially help general game players in endgame evaluation.

The remainder of the paper is organized as follows. Section~2 reviews preliminaries on ASP, GDL, and XSSAT. Section~3 introduces SQASP and presents the translation from GDL to SQASP. Section~4 evaluates the proposed approach on a set of stochastic games. We conclude in Section~5.

%% file: background.tex
\section{Preliminaries}
We review the basics of Answer Set Programming (ASP)~\cite{gebser2012answer}, Game Description Language (GDL)~\cite{genesereth2005general}, and Extended Stochastic Satisfiability (XSSAT)~\cite{teige2012stochastic}. 
\subsection{Answer Set Programming}
We consider normal logic programs with normal rules, choice rules, and integrity constraints. A rule $r$ has the form $H \aif a_{1},\ldots,a_{m},~not~a_{m+1},\ldots,~not~a_{n}.$, where $H$ is either an atom $p$, in which case we call $r$ a normal rule; or $\{p\}$, making r a choice rule; or empty, so that $r$ an integrity constraint. Rules with variables are viewed as an abbreviation for the set of their ground instances. Semantically, we identify the head of a choice rule $\{p\}$ with $p \vee \neg p$, the head of an integrity constraint with $\bot$, any rule with $(a_{1} \wedge \ldots \wedge a_{m} \wedge \neg a_{m+1} \wedge \ldots \wedge \neg a_{n}) \rightarrow H$, and a program with the conjunction of its rules. A set of atoms $\mathcal{X} \subseteq \mathcal{A}$ is a \emph{stable model\/}/\emph{answer set\/} of a program $P$ over ground atoms $\mathcal{A}$ if it is a subset-minimal model of the formula that results from replacing in $P$ any literal by $\bot$ if it is not satisfied by $\mathcal{X}$. $SM(P)$ denotes the set of answer sets of $P$.
\subsection{Game Description Language}
\label{sec:gdl}
\begin{figure}[t]
    \footnotesize
    \begin{verbatim}
    role(x). role(o). role(random). init(step(0)).
    base(step(0)). base(win).  input(x,le). input(x,ri).
    input(o,noop). input(random,a). input(random,b). input(random,c).  
    legal(R,A):- input(R,A). 
    next(win):- does(x,le), does(random, a).
    next(win):- does(x,le), does(random, b).
    next(win):- does(x,ri), does(random, c).  
    terminal:- not true(step(0)).  goal(x,100):- true(win).
    \end{verbatim}
\caption{GDL description of a stochastic game with horizon 1. Player $x$ gets 100 points if $win$ holds. At step 0, $x$ may choose either $le$ or $ri$, while $o$ can only play $noop$. The player $random$ chooses $a$, $b$, or $c$ uniformly, i.e.\ each with probability $\frac{1}{3}$. Hence, if $x$ selects $le$ (resp. $ri$), the probability that $win$ holds in the next state is $\frac{2}{3}$ (resp. $\frac{1}{3}$).}
\label{fig:gdl}
\end{figure}
The Game Description Language (GDL) can be used to describe the rules of any finite deterministic game with complete information as a normal logic program. GDL-II, an extension of GDL, can be used to model stochastic games with imperfect information. We focus on a fragment of GDL-II (which we still denote as GDL) that describes stochastic games with complete state information. The following are the preserved keywords of this fragment that describe the different elements of a game. \texttt{random} is a special player that is used to model the stochastic environment, while the remaining keywords are from the original GDL:
\begin{center}
    \footnotesize
\begin{tabular}{|l|l||l|l|}
    \hline
    \texttt{role(P)} \ &\ \texttt{P} is a player \ &\
      \texttt{input(P,M)} \ &\ \texttt{M} is in the move domain of \texttt{P}\\ 
    \texttt{random} \ &\ The random player \ &\
      \texttt{legal(P,M)} \ &\ \texttt{M} is a valid action for \texttt{P}\\
    \texttt{base(F)} \ &\ \texttt{F} is a base proposition \ &\
      \texttt{does(P,M)} \ &\ Player \texttt{P} performs action \texttt{M}\\
    \texttt{init(F)} \ &\ \texttt{F} holds in the initial state \ &\ 
      \texttt{terminal} \ &\ The current state is terminal\\
    \texttt{true(F)} \ &\ \texttt{F} holds in current state \ &\
      \texttt{goal(P,V)} \ &\ \texttt{P} gets point \texttt{V} in current state\\
    \texttt{next(F)} \ &\ \texttt{F} holds in the next state \ &\ \ &\ \\
        \hline
\end{tabular}
\end{center}
There are further restrictions for a set of GDL rules to be \emph{valid\/}~\cite{genesereth2005general}. For example, 
\texttt{role} can appear only in facts; 
\texttt{init} and \texttt{next} can only appear as heads of rules; and \texttt{true} and \texttt{does} only appear in rule bodies; 
Moreover, \texttt{legal}, \texttt{terminal}, and \texttt{goal} do not depend on \texttt{does}. Finally, a GDL description must be well-formed, allowed, and stratified--such logic programs admit a unique stable model. As a very simple example, Fig.~\ref{fig:gdl} shows the GDL description of a one-shot game.


A valid GDL description $G$ over ground terms~$\Sigma$ can be interpreted as a multi-agent transition system\/: Let $\beta=\{f\in\Sigma\, |\, G \models \gdl{base}(f)\}$ be the \emph{base propositions}, and $\gamma(p)=\{a \mid G \models \gdl{input}(p,a)\}$ be the \emph{\em move domain of $p$}. Suppose that $S=\{f_{1},\myDots,f_{n}\} \subseteq \beta$ is any state and $A=\{p_{1},\myDots,p_{k}\} \rightarrow \Sigma$ be any function that assigns to each of $k\geq 1$ players an action from $\gamma(p_{i})$. To use the game rules~$G$ to determine the state update, $S$~needs to be encoded as a set of facts using keyword $\gdl{true}$\/: $S^{true}=\{\gdl{true}(f_{1}).,\myDots,~\gdl{true}(f_{n}).\}$ and the joint action~$A$ by a set of facts using keyword $\gdl{does}$\/: $A^{does}=\{\gdl{does}(p_{1},A(p_{1})).,~\myDots,~\gdl{does}(p_{k},A(p_{k})).\}$. 

The \emph{semantics of a GDL description~$G$\/}~\cite{schiffel2010multiagent} is the state transition system\/:
\begin{itemize}
\item $R=\{p \in\Sigma \,|\, G \models\gdl{role}(p)\}$ (player names) 

\item $\Sinit=\{f\in\beta\,|\,G \models\gdl{init}(f)\}$ (initial state) 

\item $T=\{S\subseteq\beta\,|\,G \cup S^{true}\models\gdl{terminal}\}$  (terminal states)

\item $l=\{(p, a, S)\,|\, G \cup S^{true}\models\gdl{legal}(p,a)\}$ (legal moves) 

\item $u(A, S)\! =\! \{f \in \beta\,|\, G \cup S^{true} \cup A^{does}\! \models\! \gdl{next}(f)\}\!$ (update) 

\item $g = \{(p, v, S) \,|\, G \cup S^{true} \models \gdl{goal}(p,v)\}$ (goal value)
\end{itemize}
Let $lg(p,S)=\{a \mid (p,a,S) \in l\}$ denote the set of legal actions of player $p$ at position $S$. E.g., in the game in Fig.~\ref{fig:gdl} we have $lg(x,\{step(0)\})=\{le,ri\}$. 

A GDL game proceeds as follows: starting with $S=\Sinit$, at each $S$ each player $p \in R$ selects one of its moves $m$ such that $m \in lg(p,S)$, where the player \gdl{random} chooses move uniformly so that each legal action has a probability of $\frac{1}{|lg(random,S)|}$ of being selected. The joint move $A$ is then applied to the current state, resulting in the next state $u(A,S)$. We represent a \emph{valid sequence\/} of $n$ steps as $S_{0}~\xrightarrow[]{A_{0}}~S_{1}~\xrightarrow[]{A_{1}}~...~\xrightarrow[]{A_{n-1}}~S_{n}$
where $S_{i}\notin T$ and \ $(p, A(p), S_{i}) \in l$ for all $i<n$ and $p\in R$. For example, $S_0=\{step(0)\}\xrightarrow[]{\langle le,noop,b\rangle}\{win\}=S_1$ in the game in Fig.~\ref{fig:gdl}, with $S_1\in T$ a terminal state.

A sequence \emph{terminates in $n$ steps\/} if $S_{n} \in T$. Since all well-formed GDL descriptions must be terminating, we define the \emph{horizon} of a game as the minimum $n$ such that all valid sequences terminate within $n$ steps. 

Due to the similarities of syntax and semantics, ASP is a natural choice for reasoning about GDL games up to a bounded depth. To this end, an ASP program can be created by extending GDL rules by a ``time'' dimension~\cite{thielscher2010temporal}. 
Given a valid GDL $G$ description, its Temporal-Extension with horizon $n\geq 0$ is defined as $\pext^{n}(G)=\bigcup_{0 \leq i \leq n} \{c^{i}\mid c \in G\}$ where $\cdot^{i}$ replaces each occurrence of
\begin{itemize}
    \item  $\gdl{init}(f)$ by $\gdl{true}(f,0)$; and $\gdl{next}(f)$ by $\gdl{true}(f,i+1)$.
    \item $\gdl{q}(\vec{t}\,)$ by $\gdl{q}(\vec{t},i)$ if $q \notin \{init,next\}$ is a predicate symbol with argument $\vec{t}$.
\end{itemize}
We further define $A^{does}(i)= \{\gdl{does}(r_{1},A(r_{1}),i),\myDots, \gdl{does}(r_{k}, A(r_{k}),i)\}$.

$\pext^{n}(G)$ can be viewed as $\nmax+1$ copies of $G$, hence, it is stratified whenever $G$ is. Theorem~\ref{tem:semantic} shows the semantics equivalence between $\pext^{n}(G)$ and $G$.
\begin{theorem}[\cite{thielscher2010temporal}]
\label{tem:semantic}
Given a GDL description $G$ and a valid sequence $\Sinit \xrightarrow[]{A_{0}} S_{1} \xrightarrow[]{A_{1}} \myDots S_{m-1} \xrightarrow[]{A_{m-1}} S_{m}$. Let $P=\pext^{n}(G) \cup A_{0}^{does}(0)\cup \ldots \cup~A_{m-1}^{does}(m-1)$ for some $n \geq m$. Then, for any predicate symbol $p$ in the game description $G$ and for all $0\leq i\leq m$, such that $p \notin \{\texttt{init},\texttt{next}\}$ and $p$ does not depend on $\texttt{does}$ in $G$, we have 1) $S_{i}=\{f \mid P \models true(f, i)\}$, and 2) $G \cup S_{i}^{true} \models p(t)$ iff $P \models p(t,i)$. 
\end{theorem}
\subsection{Extended Stochastic Satisfiability}
\label{sec:XSSAT}
Extended Stochastic Satisfiability (XSSAT)~\cite{teige2012stochastic} extends QBF with chance quantifiers ($\rotR$). An XSSAT formula over variables $\mathcal{V}$ is of the form $\Phi=\mathcal{Q}. \psi$ where $\mathcal{Q}=Q_{1}. v_{1},Q_{2}. v_{2},\ldots,Q_{n}. v_{n}$ is the quantifier prefix and $\Phi$ is a propositional formula in conjunctive normal form (CNF). Each $Q_{i} \in \{\exists,\forall,\rotR^{p_{i}}\}$ and $v_{i} \in \mathcal{V}$ where $\rotR^{p_{i}} v_{i}$ means that the boolean variable $v_{i}$ has a probability of $p_{i}$ to be assigned to be $\top$. The semantics of $\Phi$ is defined as the maximal probability of satisfying $\Phi$, which is given recursively as follows, where $\epsilon$ is the empty prefix:
\begin{itemize}
    \item $Pr[\epsilon. \psi]=1$ if $\psi$ is satisfiable and $0$ if $\psi$ is unsatisfiable
    \item $Pr[\exists x~\mathcal{Q}.\psi]=max(\mathcal{Q}. (\psi \wedge x),\mathcal{Q}. (\psi \wedge \neg x))$.
    \item $Pr[\forall x~\mathcal{Q}.\psi]=min(\mathcal{Q}. (\psi \wedge x),\mathcal{Q}. (\psi \wedge \neg x))$.
    \item $Pr[\rotR^{p} x~\mathcal{Q}.\psi]=p \cdot Pr[\mathcal{Q}. (\psi \wedge x)] + (1-p) \cdot Pr[\mathcal{Q}. (\psi \wedge \neg x)]$.
\end{itemize}

%% file: encoding.tex
\section{Solving Two-Player Stochastic GDL Games}
We present an ASP-based approach to solving two-player stochastic GDL games for a finite horizon. W.l.o.g., we denote the two adversarial players as $x$ and $o$ respectively. We also abbreviate the $\gdl{random}$ player as $\gdl{r}$. We first formally define two-player turn-taking stochastic games and the maximal winning probability of player $x$ within $\nmax$ steps in the context of GDL. 
\begin{definition}
\label{def:2playergame}
    A GDL game $G$ with semantics $(R=\{x,o,\random\},\Sinit,T,l,u,g)$ is a two-player turn-taking stochastic game iff for any non-terminal game position that is reachable from $\Sinit$, at most one player in $\{x,o\}$ has more than one legal action. The maximal probability of player $x$ to win the game $G$ within $n$ steps starting at state $S$ (denoted as $P_{xw}(S,n)$) is recursively defined as follows:
    \begin{enumerate}
        \item If $S \in T$ and $n \geq 0$ then $P_{xw}(S,n)=1$ if $(x,100,S) \in g$; else $P_{xw}(S,n)=0$.
        \item If $S \notin T$ and $n = 0$ then $P_{xw}(S,n)=0$
        \item Otherwise (below $A=\{(x,a_{x}),(o,a_{o}),(\random,a_{r})\}$),   
        \[P_{xw}(S,n)=\max_{a_{x} \in lg(x,S)}(\min_{a_{o} \in lg(o,S)}(\frac{1}{|lg(\random,S)|} \sum_{a_{r} \in lg(\random,S)} P_{xw}(u(A,S),n-1))) \] 
    \end{enumerate}
\end{definition}
Our definition provides a pessimistic estimation of the winning probability of player $x$, assuming that player $o$ always tries to minimize such a probability. However, if $G$ is also a zero-sum two-outcome game, it is known that minimizing the winning probability of $x$ is the same as player~$o$ maximizing its own. For the game in Fig.~\ref{fig:gdl}, $P_{xw}(\Sinit,1)=0.67$, which can be achieved by $x$ playing $le$ at $\Sinit$.
\subsection{Stochastic Quantified Answer Set Programming}
Due to the existing ASP-based approach to solving single- or two-player GDL games~\cite{he2024solving,thielscher2009answer}, using an ASP-based approach to solve stochastic games is a natural idea. However, none of the existing ASP frameworks can express both the adversarial and stochastic features of such games. We define a new language, Stochastic Quantified Answer Set Programming (SQASP), that generalizes QASP~\cite{fandinno2021planning}. 
\begin{definition}[SQASP]
\label{def:sqasp}
Suppose $P$ is a normal logic program over ground atoms $\mathcal{A}$. A \emph{stochastic quantified answer set program\/} over $\mathcal{A}$ has the form $Q_{1}. x_{1}~\ldots~Q_{n}. x_{n}~~P$, where $Q_{i} \in \{\exists,\forall,\rotR\}$, $x_{i} \in \mathcal{A}$, and $x_{i} \neq x_{j}$ for any $i \neq j$. The maximal probability of satisfying an SQASP is recursively defined as: 
\begin{enumerate}
    \item $Pr[\epsilon.P]=1$ if $P$ has a stable model and $0$ if $P$ has no stable models. 
    \item $Pr[\exists x. \mathcal{Q}. P]=max(\mathcal{Q}. P \cup \{\aif~x.\},\mathcal{Q}. P \cup \{\aif~not~x.\})$.
    \item $Pr[\forall x. \mathcal{Q}. P]=min(\mathcal{Q}. P \cup \{\aif~x.\},\mathcal{Q}. P \cup \{\aif~not~x.\})$.
    \item $Pr[\rotR^{p} x. \mathcal{Q}. P]=p \cdot Pr[\mathcal{Q}. P \cup \{\aif~not~x.\}] + (1-p) \cdot Pr[\mathcal{Q}. P \cup \{\aif~x.\})]$.
\end{enumerate}
\end{definition}
Similar to deciding QASP with $qasp2qbf$~\cite{fandinno2021planning}, for any SQASP $\mathcal{Q}. P$, $Pr[\mathcal{Q}. P]$ can be computed by using tools like $lp2sat$~\cite{janhunen2011compact} to translate $P$ into a CNF $\psi$ whose models coincide with the stable models of $P$, and then evaluating the XSSAT formula $\mathcal{Q}. \psi$. The correctness of this approach is established by the following theorem, which can be proved by induction on the number of variables in $\mathcal{Q}$. 
\begin{theorem}
\label{theorem:sqasp2XSSAT}
    Given a normal logic program $P$ over ground atoms $\mathcal{A}$ and a CNF $\psi$ over $\mathcal{A} \cup \mathcal{B}$ such that $SM(P)=\{M \cap \mathcal{A}\mid M\text{~is a model of~}\psi)\}$. Then for any quantifier prefix $\mathcal{Q}=Q_{1}. x_{1} \ldots Q_{n}. x_{n}$ such that for all $1 \leq i \leq n$, $x_{i} \in \mathcal{A}$ and $Q_{i} \in \{\exists,\forall,\rotR^{p}\}$, $Pr[\mathcal{Q}.~P]=Pr[\mathcal{Q}.~\psi]$.
\end{theorem}
We implement a tool called $sqasp2xssat$ by extending $qasp2qbf$~\cite{fandinno2021planning}, which converts any SQASP program with rational probability into an XSSAT formula with the same maximal satisfiability probability. The tool takes a program $\mathcal{Q}.P$ as input, where $\mathcal{Q}$ is specified as facts over the reserved predicates $\_exists/2$, $\_forall/2$, and $\_chance/4$. In particular, the new predicate $\_chance(i,p,q,a)$ says that $a$ is quantified at level $i$ with probability $\frac{p}{q}$ of being forced to true. 
With this new tool, we can solve any stochastic two-player game using XSSAT solvers, as long as it can be encoded in SQASP\footnote{The reason that we do not translate GDL to XSSAT directly is that GDL is based on stable model semantics, while XSSAT uses classical models. A direct translation is equivalent to implementing $lp2sat$~\cite{janhunen2011compact} from scratch, which is tedious.} 
\subsection{Transforming GDL Games into SQASP}
Our goal is to construct an SQASP program $\Psi$ such that $Pr[\Psi]=P_{xw}(\Sinit,\nmax)$. The encoding proceeds in two steps. First, we view the game as cooperative between players $x$, $o$, and $\random$, and construct a logic program $P_{c}^{\nmax}$ such that $\pext^{\nmax}(G)\cup P_{c}^{\nmax}$ ensures: a) each player executes exactly one legal action at each timestamp $0\le t<\nmax$ unless the game has terminated; b) the game terminates within $\nmax$ steps; and c) upon termination, player $x$ wins (i.e., receives a score of $100$).
Second, we extend $\pext^{\nmax}(G)\cup P_{c}^{\nmax}$ to capture adversarial and stochastic behaviors.

For the first step, we introduce an auxiliary predicate $\texttt{ended}(t)$, indicating that the game has terminated by time~$t$. Let $P_{c}^{\nmax}=\bigcup_{t=0}^{\nmax} P_{c}(t)$, where each $P_{c}(t)$ consists of rules~\ref{eq:1}--\ref{eq:6}. 
\begin{enumerate}[label=(\arabic{enumi}):~,ref=(\arabic{enumi}),align=right,leftmargin=\widthof{(100)}+\labelsep]
    \item \label{eq:1} $1~\{does(p,A,t) : input(p,A) \}~1~\aif~not~ended(t).$ (for $t<\nmax$ only)
    \item \label{eq:2} $\aif~not~legal(p,A,t),~does(p,A,t).$ (for $t<\nmax$ only)
    \item \label{eq:3} $ended(t)~\aif~terminal(t).$
    \item \label{eq:4} $ended(t)~\aif~ended(t-1).$ (for $t>0$ only)
    \item \label{eq:5} $\aif~not~ended(\nmax).$
    \item \label{eq:6} $\aif~not~goal(x,100,t),~ended(t),~not~ended(t-1).$
\end{enumerate}
Rules~\ref{eq:1}--\ref{eq:2} model condition~a), rule~\ref{eq:3}--\ref{eq:5} model condition~b), and~rule \ref{eq:6} ensures that condition~c) is satisfied.


In the second step, we encode adversarial and stochastic choices of $o$ and $\random$. For each player $p\in \{o,\random\}$, we fix an arbitrary ordering from $1$ to $|\gamma(p)|$ of actions in $\gamma(p)$ and let $ord(p,a)$ denote the position of action $a$.

To model player~$o$, we adapt the corrective encoding used for deterministic two-player games~\cite{he2024solving}. 
We define $P_{o}^{\nmax}=\bigcup_{t=0}^{\nmax-1} \bigcup_{a \in \gamma(o)} P_{o}(a,t)$ where for each $0 \leq t <\nmax$ and $a \in \gamma(o)$, $P_{o}(a,t)$ consists of rules~\ref{eq:7}–\ref{eq:8}:
\begin{enumerate}[label=(\arabic{enumi}):~,ref=(\arabic{enumi}),align=right,leftmargin=\widthof{(100)}+\labelsep]
\setcounter{enumi}{6}
    \item \label{eq:7} $\{dec\_o(L,t)\}~\aif~ldom(L).$
    \item \label{eq:8} $\aif~not~does(o,a,t),~legal(o,a,t),~not~ended(t),$
    \item[] $\hspace{1em}~~dec\_o(\rho_{1},t),\myDots,dec\_o(\rho_{j},t),not~dec\_o(\mu_{1},t),~\myDots,~not~dec\_o(\mu_{k},t).$
\end{enumerate}
where $\rho_{1},\ldots,\rho_{j}$ are the 1 bits in the binary representation of $ord(o,a)-1$, and $\mu_{1},\ldots,\mu_{k}$ are the 0 bits in the binary representation of $ord(o,a)-1$.


The predicate $\texttt{dec\_o}$ encodes actions logarithmically in which the domain of $\texttt{ldom}$ ranges from $1$ to $\lceil\log_{2} |\gamma(o)| \rceil$.  A rule of the form~\ref{eq:8} is introduced, resulting in $|\gamma(o)|$ rules for each timestamp $t$. 
Universal quantification over $\texttt{dec\_o}$ captures adversarial feature of the game: at time~$t$, player~$o$ universally selects a binary combination of $\texttt{dec\_o}$. If the binary combination of $\texttt{dec\_o}$ evaluates to be $i$ and the $(i+1)$-th action in the ordering is legal for player $o$ at timestamp $t$,~\ref{eq:8} forces~\ref{eq:1} to generate the corresponding action.  


To model the stochastic feature of the game, we must ensure that, if the game does not terminate at timestamp~$t$, player~$\random$ selects exactly one legal action according to a \emph{uniform distribution}. The encoding consists of two parts. 

First, we construct a program $P_{legal}^{\nmax}=\bigcup_{t=0}^{\nmax-1} P_{legal}^{\nmax}(t)$ where each $P_{legal}^{\nmax}(t)=\{\ref{eq:9},\ref{eq:10},\ldots,\ref{eq:14}\}$ extracts the set of legal actions for player~$\random$ at timestamp~$t$ and imposes an ordering on these actions. Next, since at each timestamp~$t$ before termination player~$\random$ may have between $1$ and $|\gamma(\random)|$ legal actions, we construct a program $P_{r}^{\nmax}=\bigcup_{t=0}^{\nmax-1} \bigcup_{i=2}^{|\gamma(\random)|} P_{r}(i,t)$ where each $P_{r}(i,t)=\{\ref{eq:15},\ref{eq:16}\}$ ensures that every legal action extracted by $P_{legal}$ is selected with probability $\frac{1}{i}$ when the total number of legal actions available to player~$\random$ at timestamp~$t$ is~$i$.
\begin{enumerate}[label=(\arabic{enumi}):~,ref=(\arabic{enumi}),align=right,leftmargin=\widthof{(100)}+\labelsep]
\setcounter{enumi}{8}
    \item \label{eq:9} $succ(a_{r}^{1},a_{r}^{2}). \ \ succ(a_{r}^{2},a_{r}^{3}). \ \ \ldots \ \ succ(a_{r}^{|\gamma(\random)|-1},a_{r}^{|\gamma(\random)|}).$
    \item \label{eq:10} $count(a_{r}^{1},0,t)~\aif~not~legal(\random,a_{r}^{1},t).$
    \item \label{eq:11} $count(a_{r}^{1},1,t)~\aif~legal(\random,a_{r}^{1},t).$
    \item \label{eq:12} $count(B,N,t)~\aif~succ(A,B),~not~legal(\random,B,t),~count(A,N,t).$
    \item \label{eq:13} $count(B,N,t)~\aif~count(A,N-1,t),~succ(A,B),~legal(\random,B,t),~cdom(N).$
    \item \label{eq:14} $tol(N,t)~\aif~count(a_{r}^{|\gamma(\random)|},N,t).$    
        \item \label{eq:15} $\{dec\_r(1..|\gamma(\random)|,t)\}.$
    \item \label{eq:16} For each $1 \leq j \leq i$, we create the rule:
    
    \item[] $\aif~not~does(\random,A,t),~legal(\random,A,t),~count(A,j,t),~not~ended(t),~tol(i,t),$
    \item[] $\hspace{9.0em} not~dec\_r(i,t),\myDots,~not~dec\_r(j+1,t),~dec\_r(j,t).$

\end{enumerate}
Recall that we impose an ordering on the actions of player~$\random$. In $P_{legal}(t)$, we introduce an auxiliary predicate $\texttt{succ(A,B)}$ to represent this ordering, where $B$ is the successor of $A$, and a predicate $\texttt{count(A,N,t)}$ indicating that at time~$t$ there are $N$ legal actions whose ordering does not exceed~$ord(\random,A)$. The value of $N$ ranges from $0$ to~$|\gamma(\random)|$. Rules~\ref{eq:10}–\ref{eq:14} inductively compute these counts. Consequently, an action $a$ is the $i$-th legal action of player~$\random$ at time~$t$ if and only if both $count(a,i,t)$ and $legal(\random,a,t)$ hold.

To model uniform random choice, we introduce in $P_{r}(i,t)$ an auxiliary predicate $\gdl{dec\_r}$ whose truth values uniquely determine the action selected by~$\random$. For each timestamp $0 \leq t < \nmax$, we introduce $|\gamma(\random)|$ instances of $\texttt{dec\_r}$ and quantify $\texttt{dec\_r}(i,t)$ using the chance quantifier $\rotR^{\frac{1}{i}}$. The program $P_{r}(i,t)$ applies when player~$\random$ has exactly $i$ legal actions at time~$t$ and ensures that the selected action is uniformly distributed. In particular, the $j$-th rule of~\ref{eq:16} says that the $j$-th legal action of $\random$ must be generated if $\texttt{dec\_r}(j,t)$ holds and no $\texttt{dec\_r}(k,t)$ with $j<k\leq i$ holds. By construction, each legal action is selected with probability $\frac{1}{j} \cdot (1-\frac{1}{j+1}) \cdot \ldots \cdot (1-\frac{1}{i})=\frac{1}{i}$ which satisfies the uniform distribution requirement. The case $i=1$ is handled directly by rules~\ref{eq:1} and~\ref{eq:2} in $P_{c}^{\nmax}(t)$ and therefore the program $P_{r}(1,t)$ is not required.


To sum up, our SQASP is of the form $\Psi_{b}=\mathcal{Q}_{b}~\Phi^{\nmax}$ where
\begin{center}
    $\Phi^{n}=\pext^{\nmax}(G) \cup P_{c}^{\nmax} \cup P_{o}^{\nmax} \cup P_{legal}^{\nmax} \cup  P_{r}^{\nmax}$   
\end{center}
and the quantifier prefix $\mathcal{Q}_{b}$ is obtained by quantifying all the \textbf{choice atoms} (i.e., instances of $\texttt{does}$, $dec\_o$, and $dec\_r$) according to the temporal order, while placing all the other atoms in the innermost existential quantifier block.
\begin{definition}[Baseline Quantification Method]
    \label{def:naive}
    Suppose that $G$ is a valid GDL description of a two-player stochastic turn-taking game and $\nmax$ is the maximum number of steps allowed in the game. Let $\mathcal{A}$ be the set of ground atoms of the program $\Phi^{\nmax}$. The quantifier prefix $\mathcal{Q}_{b}$ is of the form:
    \begin{center}
        $\mathcal{Q}_{b}=E_{0} U_{0} E_{1} R_{0} E_{2} U_{1} E_{3} R_{1} E_{4} \ldots U_{n-1} E_{2 \cdot n-1} R_{n-1} E_{2 \cdot n}$
    \end{center}
    For each $a \in \mathcal{A}$, the quantifier block it belongs to is determined as follows:
    \begin{enumerate}
        \item If $a=does(x,A,i)$ for some $A$ and $0 \leq i < n$, then $\exists a \in E_{2 \cdot i}$.
        \item If $a=dec\_o(L,i)$ for some $L$ and $0 \leq i < n$, then $\forall a \in U_{i}$.
        \item If $a=does(o,A,i)$ for some $A$ and $0 \leq i < n$, then $\exists a \in E_{2 \cdot i + 1}$.
        \item If $a=dec\_r(X,i)$ for some $X$ and $0 \leq i < n$, then $\rotR^{\frac{1}{X}} a \in R_{i}$.
        \item If $a=does(\random,A,i)$ for some $A$ and $0 \leq i < n$, then $\exists a \in E_{2 \cdot i + 2}$.
        
        \item Otherwise, $\exists a \in E_{2 \cdot \nmax}$.
    \end{enumerate}
\end{definition}
The following theorem establishes the correctness of the overall encoding. The proof is based on Theorem~\ref{tem:semantic} and is similar to the correctness proof of the encoding of two-player deterministic games~\cite{he2024solving}.
\begin{theorem}
    Suppose that $G$ is a valid GDL description of a two-player stochastic turn-taking game, then for any $\nmax \geq 0$, we have that $Pr[\Psi_{b}]=P_{xw}(\Sinit,\nmax)$.
\end{theorem}
\subsection{Quantifier Shifting}
In the baseline encoding, all non-choice atoms are quantified in the innermost existential block. However, for two-player deterministic games, atoms are quantified as early as possible for performance reasons~\cite{he2024solving}. For example, in any valid GDL description, the value of $\gdl{legal}$ at time~$t$ does not depend on $\gdl{does}$ at time $\geq t$, and can therefore be quantified before $\gdl{does}$ at the same timestamp. We hypothesize that such early quantification is also beneficial for stochastic games. Since $\pext^{\nmax}(G)$ is stratified and $\gdl{does}$ appears only in rule bodies, removing the choice rules~\ref{eq:1},~\ref{eq:6}, and~\ref{eq:15} from $\Phi^{\nmax}$ yields a stratified program with integrity constraints in which choice atoms occur only in bodies, the value of any non-choice atom is therefore uniquely determined once all choice atoms it depends on are fixed. This motivates the following (generalized) dependency-based quantification method that preserves the temporal order of choice atoms and quantifies each non-choice atom immediately after the choice atoms it depends on.
\begin{definition}
\label{def:dep}
    Given a GDL description $G$ and an encoding depth $\nmax$. Suppose that $P$ is the ground program of $\Phi^{\nmax}$ over ground atoms $\mathcal{A}$. An atom $p$ depends on $q$ in $P$ (denoted $p \rightarrow q$) iff $P$ has a rule such that $p$ appears in the head and $q$ appears in the body; or, there exists $x \in \mathcal{A}$ such that $p \rightarrow x$ and $x \rightarrow q$. The quantifier prefix $\mathcal{Q}_{d}$ obtained by the \textbf{dependency-based quantification method} is of the same form as $\mathcal{Q}_{b}$ in Definition~\ref{def:naive}, where for each $a \in \mathcal{A}$, the quantifier block it belongs to is determined as follows:
    \begin{itemize}
        \item If $a$ is a choice atom, the quantifier block it belongs to is determined by the first 5 rules as in~Definition~\ref{def:naive}
        \item Otherwise, $\exists a \in E_{t}$ where $t$ is the maximal $1 \leq t \leq 2 \cdot n$ such that:
        \begin{itemize}
            \item $a \rightarrow does(x,A,\frac{t}{2})$ for some $does(x,A,\frac{t}{2}) \in \mathcal{A}$; or
            \item $a \rightarrow does(o,A,\frac{t-1}{2})$ for some $does(o,A,\frac{t-1}{2}) \in \mathcal{A}$; or 
            \item $a \rightarrow does(\random,A,\frac{t-2}{2})$ for some $does(\random,A,\frac{t-2}{2}) \in \mathcal{A}$; or
           \item $a \rightarrow dec\_o(L,\frac{t-1}{2})$ for some $dec\_o(L,\frac{t-1}{2}) \in \mathcal{A}$; or 
           \item $a \rightarrow dec\_r(X,\frac{t-2}{2})$ for some $dec\_r(X,\frac{t-2}{2}) \in \mathcal{A}$. 
        \end{itemize}
        If no such $t$ exists, $\exists a \in E_{0}$.
    \end{itemize}
\end{definition}
We remark that our encoding generalizes existing encodings for single- and two-player deterministic games and can also be applied to single-player stochastic games, where the objective is to compute the maximal probability of winning a game within $\nmax$ steps against a random player. The building blocks for solving the different types of games are summarized as follows, where $\mathcal{Q} \in \{\mathcal{Q}_{b},\mathcal{Q}_{d}\}$\/:
\begin{center}
\footnotesize
\begin{tabular}{|l|l|}
    \hline
   Single-Player Deterministic Game~\cite{thielscher2009answer}  & $\pext^{\nmax}(G) \cup P_{c}^{\nmax}$ \\
   Two-Player Deterministic Game~\cite{he2024solving} & $\mathcal{Q}~\pext^{\nmax}(G) \cup P_{c}^{\nmax} \cup P_{o}^{\nmax}$ \\
   Single-Player Stochastic Game& $\mathcal{Q}~\pext^{\nmax}(G) \cup P_{c}^{\nmax} \cup P_{legal}^{\nmax} \cup P_{r}^{\nmax}$ \\
   Two-Player Stochastic Game&  $\mathcal{Q}~\pext^{\nmax}(G) \cup P_{c}^{\nmax} \cup P_{o}^{\nmax} \cup P_{legal}^{\nmax} \cup P_{r}^{\nmax}$ \\
   \hline
\end{tabular}
\end{center}

%% file: experiment.tex
\section{Experimental Results}
We evaluate our translation on both single-player and two-player stochastic games. Backgammon is a popular two-player stochastic game, but its state space is too large for perfect play, so we focus on smaller games. For single-player stochastic games, we compute the maximal winning probability against a random opponent on the two-player games Connect-3 (C‑3), Connect‑4 (C‑4), and Tic‑Tac‑Toe (TT). For two-player stochastic games, we consider stochastic board games and stochastic Nim (S‑N)~\cite{bertolino2017monte}. Stochastic board games (S-C-3, S-C-4, S-TT) are motivated by stochastic Tic-Tac-Toe~\cite{STT}, where we use the original board game rule, despite that at each round, each player chooses the cell it wants to mark while there is a probability of $p$ that the cell is marked as its own color and $1-p$ that it is marked as the color of its opponent. The Stochastic nim variant we considered operates on one pile of size $n$, and the two players alternately take 1 or 2 pieces from the top of the pile. The first player to remove all pieces from the pile wins the game. Between each turn, the random player may add 1 piece to the pile with a probability of $p$ or do nothing otherwise. To ensure that the game satisfies the terminating requirement for well-formed game descriptions, we define it to end in a draw after each player makes $n$ moves. 

Our implementation takes a GDL description and depth as input and creates an SQASP program using either the baseline ($\mathcal{Q}_{b}$) or the dependency-based quantification method ($\mathcal{Q}_{d}$). For simplicity, the depth of each instance is set to be the horizon of the game based on human domain knowledge. For board games, the horizon is obviously equal to the size of the board, and for stochastic nim, the horizon of each game is $4 \cdot n$ where $n$ is the initial pile size. Then, the SQASP program is converted to an XSSAT instance using $sqasp2xssat$ and solved with the only publicly available XSSAT solver \emph{SharpSSAT}~\cite{fan2023sharpssat}. For comparison purposes, we also implement an expected-minimax solver with transposition tables (\emph{EMM}) that uses Prolog as the GDL reasoner for legal actions. All games were run on a laptop with an operating system 64-bit Ubuntu 24.04, 2.6 GHz CPU, and a timeout of 20 minutes per game\footnote{Link to the prototype and experiments: \url{https://github.com/hharryyf/gdl2ssat/}}. 
The following table summarizes the configuration of each game, the maximal achievable winning probability for both players, and the runtime of each method (in seconds) for computing these probabilities.\footnote{* means that the solver timed out after 1200 seconds.}
\begin{center}
    \footnotesize
    \begin{tabular}{|c|r|r|r|r|r|r|r|r|r|r|}
    \cline{1-11}
        \textbf{Game} & \textbf{Config} & $p$ & \multicolumn{4}{c|}{\textbf{First}} &  \multicolumn{4}{c|}{\textbf{Second}}  \\
        \cline{4-11}
        & & &
       $\%$ & \textbf{$\mathcal{Q}_{b}$}  & \textbf{$\mathcal{Q}_{d}$} & \textbf{EMM}  
       &  \multicolumn{1}{r|}{$\%$} & \multicolumn{1}{r|}{\textbf{$\mathcal{Q}_{b}$}}  & \multicolumn{1}{r|}{\textbf{$\mathcal{Q}_{d}$}} & \multicolumn{1}{r|}{\textbf{EMM}}  \\
        \cline{1-11}
        & 4x4 & n/a& 100.00 &  \textbf{0.31}& 0.47 & 1.05 & 98.26  & \textbf{2.17} & 2.19 &8.01\\
        \cline{2-11}
       C-3 & 5x5 &n/a & 100.00 & 149.51&  \textbf{19.29}& 336.48 & 99.20 &  \textbf{773.59} & 876.28 &  * \\
       \cline{2-11}
         & 6x6 & n/a& 100.00& *& * &*  &?& *& * &*  \\
       \cline{1-11}
         & 4x4& n/a& 95.77& \textbf{4.16}& 4.28 & 58.68 & 96.69 &  4.08 & \textbf{3.85} & 61.92 \\
         \cline{2-11}
       C-4 & 5x4& n/a& 99.76& \textbf{264.04}&  451.35& * & 99.96 &  \textbf{244.84} & 275.57 & * \\
       \cline{2-11}
         & 5x5& n/a& ?& *& * &*  & ?& *& * &* \\
          \hline
        TT & 3x3 & n/a& 99.48 & 0.70& \textbf{0.68} & 1.57 & 93.86 &\textbf{0.80} & 0.82&  1.99 \\
        \cline{2-11}
        & 4x4 & n/a& 100.00& \textbf{283.91}& 351.03 & * & ?& *& * &*\\
        \hline
       S-C-3 & 4x4  & 0.5 &  50.10& 47.83& \textbf{46.13} & 103.43  &    49.87 & \textbf{47.65} & 48.08 & 109.98   \\
       \cline{3-11}
       & & 0.8 & 60.75& 151.84 & \textbf{149.57} & 205.36 & 39.24 &  \textbf{143.62} & 155.03 & 246.24 \\
       \cline{2-11}
            & 5x5 & 0.8 &  ?& *& * &*  &  ? & * & * & *  \\
        \hline
       S-C-4 
       & 4x4 & 0.5 & 37.20& 63.29& \textbf{62.90} & *  &  37.34 & 64.28 & \textbf{63.51} & *  \\
       \cline{3-11}
       & & 0.8 & 26.26& 179.21 & \textbf{178.02} & * &  42.87 & \textbf{186.42}  & 200.63 & *  \\
            
       \cline{2-11}
            & 5x4 & 0.8 & ?& *& * &*  & ? &  * & * & *  \\
       \hline
       S-TT & 3x3& 0.5 & 47.46& 4.21& \textbf{4.06} & 10.65 &  46.29 &  4.33 & \textbf{4.02}& 11.65  \\
       \cline{3-11}
       & & 0.8 & 42.20& \textbf{9.04}& 9.69 & 23.15 &  36.74 &  10.56 & \textbf{9.90} & 22.58  \\
               
       \cline{2-11}
            & 4x4 & 0.8 & ?& *& * &*  &  ?& *& * &*    \\
         \hline
       S-N     & 30& 0.5& 50.00& 1.78& 1.89 &  \textbf{0.50}&  50.00 & 342.13  & 414.02 & \textbf{0.50}  \\
            \cline{2-11}
        & 100& 0.5& 50.00& 290.22&  292.58&  \textbf{7.18}&  50.00 &  * & * & \textbf{7.33}  \\
        \hline
    \end{tabular}
\end{center}
For board games, the configuration refers to the width and height of the board. For nim, it refers to the initial pile height. For two-player stochastic games, the value of $p$ we mentioned above is also reported. Our results reveal several interesting points\/: 
For example, in Tic-Tac-Toe, the maximal winning probability of player~$x$ exceeds $99\%$ against a random player. Moreover, in S-N, the winning probability for both players converges to $50\%$ as the initial pile size gets larger.

The two performance-related questions are\/: Is quantifier shifting as important for stochastic games as for deterministic games? Can our ASP-based approach match forward search in solving stochastic games?

We first discuss the effect of quantifier shifting. When solving two-player deterministic games, quantifier shifting is critical for the performance of the state-of-the-art QBF solver \emph{Caqe}~\cite{rabe2015caqe}. The motivation of such a technique is to reduce the search space by forcing the solvers to fix the values of atoms such as \emph{legal} or \emph{terminal} at timestamp~$t$ before \emph{does} at timestamp $t$, thereby preventing it from branching on \emph{does} atoms corresponding to illegal actions. 
However, from our results
it is surprising that the performance of \emph{SharpSSAT} under both quantification methods is similar in almost all instances. We suspect that this is because \emph{SharpSSAT} is a search-based solver~\cite{fan2023sharpssat} that has an important mechanism called \emph{unit propagation}, which allows fixing the value of certain variables without following the quantification order. Hence, although in the baseline encoding all ground atoms of \emph{legal} are quantified after \emph{does}, their values may still be derived by unit propagation once the values of \emph{does} at earlier timestamps have been fixed. It is worth noting that this result is inline with the performance of the search-based solver \emph{DepQBF}~\cite{lonsing2017depqbf} on deterministic games, where quantifier shifting is also not beneficial. Consequently, an interesting open question is whether quantifier shifting can have a more significant impact on non-search-based XSSAT solvers, should such solvers be developed in the future.

We then compare our translation with the expected minimax algorithm. Our method outperforms \emph{EMM} on all games except S-N. S-N admits a simple dynamic programming solution in which each state is defined by the step number and the pile height; expected minimax with a transposition table coincides with this strategy and therefore scales well even when the initial pile height is 100. In contrast, the XSSAT formula for S-N contains over one million clauses, which is significantly larger than the XXSAT formulas that current solvers can handle efficiently. We also note that \emph{SharpSSAT} is still under development and currently relies on naive chronological backtracking for universal variables. Its performance could be improved by integrating QBF techniques such as solution-driven cube learning~\cite{lonsing2017depqbf}. Overall, while our ASP-based approach does not dominate forward search in all cases, its performance is comparable to search-based game-solving methods and thus provides an alternative for endgame search in GGP systems. This generalizes the results for deterministic games~\cite{he2024solving,thielscher2009answer} to stochastic games.

%% file: conclusion.tex
\section{Conclusion}
Due to the similarity in syntax and semantics, ASP and its extensions have previously been applied to solving single- and two-player deterministic turn-taking games~\cite{he2024solving,thielscher2009answer}. In this paper, we extended these techniques to stochastic games by introducing the modeling language SQASP and a translation-based solver that converts SQASP programs to XSSAT and solves them using \emph{SharpSSAT}. We also studied the impact of quantifier shifting in the stochastic setting. Our experimental results show that the proposed ASP-based approach is comparable with forward search and can support endgame evaluation in general game playing. One limitation of our work is that it applies only to perfect-information games; extending it to stochastic games with partial observability, such as Poker, Krieg-Tic-Tac-Toe, or Monty Hall, is an important direction for future work.